\documentclass[aps,prc,twocolumn,showpacs,amsmath,preprintnumbers,amssymb,groupedaddress,superscriptaddress]{revtex4}
\usepackage{graphicx}
\usepackage{dcolumn}
\usepackage{bm}
\usepackage{multirow}
\usepackage{hhline}
\usepackage{bbm}

\usepackage{CJK}

\begin{document}
\begin{CJK*}{GBK}{song}

\title{Nucleon-nucleon momentum correlation function as a
probe of the density distribution of valence neutron in neutron-rich
nucleus}

\author{X. G. Cao}
\affiliation{Shanghai Institute of Applied Physics, Chinese Academy
of Sciences, Shanghai 201800, China}

\author{X. Z. Cai\footnote{E-mail address: caixz@sinap.ac.cn}}
\affiliation{Shanghai Institute of Applied Physics, Chinese
Academy of Sciences, Shanghai 201800, China}

\author{Y. G. Ma\footnote{E-mail address: ygma@sinap.ac.cn}}
\affiliation{Shanghai Institute of Applied Physics, Chinese
Academy of Sciences, Shanghai 201800, China}

\author{D. Q. Fang}
\affiliation{Shanghai Institute of Applied Physics, Chinese Academy
of Sciences, Shanghai 201800, China}

\author{G. Q. Zhang}
\affiliation{Shanghai Institute of Applied Physics, Chinese Academy
of Sciences, Shanghai 201800, China} \affiliation{Graduate School of
the Chinese Academy of Sciences, Beijing 100049, China}

\author{W. Guo}
\affiliation{Shanghai Institute of Applied Physics, Chinese Academy
of Sciences, Shanghai 201800, China}

\author{J. G. Chen}
\affiliation{Shanghai Institute of Applied Physics, Chinese Academy
of Sciences, Shanghai 201800, China}

\author{J. S. Wang}
\affiliation{Institute of Modern Physics, Chinese Academy of Sciences, Lanzhou 730000, China}
\date{\today}

\begin{abstract}
Proton-neutron, neutron-neutron and proton-proton momentum correlation functions
($C_{pn}$, $C_{nn}$, $C_{pp}$) are systematically investigated for $^{15}$C and other C isotopes induced collisions at different entrance channel conditions within the framework of the isospin-dependent quantum molecular dynamics
(IDQMD) model complemented by the CRAB (correlation after burner) computation code.
$^{15}$C is a prime exotic nucleus candidate due to the weakly bound valence neutron coupling with closed-neutron shell nucleus $^{14}$C.
In order to study density dependence of correlation function by removing the isospin effect, the initialized $^{15}$C projectiles are sampled from two kinds of density distribution from RMF model, in which the valence neutron of $^{15}$C is populated on both 1$d$5/2 and 2$s$1/2 states, respectively.
The results show that the density distributions of valence neutron significantly influence nucleon-nucleon momentum correlation function at large impact parameter and high incident energy.
The extended density distribution of valence neutron largely weakens the strength of correlation function. The size of emission source is extracted by fitting correlation function using Gaussian source method.
The emission source size as well as the size of final state phase space is larger for projectiles sampling from more extended density distribution of valence neutron corresponding 2$s$1/2 state in RMF model.
Therefore momentum correlation function can be considered as a potential valuable tool to diagnose the exotic nuclear structure such as skin and halo.
\end{abstract}
\pacs{25.60.-t, 25.70.Pq, 21.10.Gv\\}
\maketitle

\section{Introduction}
Intensity interferometry method, developed by Hanbury Brown and
Twiss (HBT) in 1950s \cite{HBT-Nature-178-1046}, was originally used to measure astronomical objects such as angular diameter of stars.
The method was later introduced into subatomic physics by
Goldhaber \emph{et al}., who extracted the spatial extent of an
annihilation fireball in proton-antiproton reactions by two-pion
correlations \cite{GG-PR-120-300}. Then the method was widely
applied in exploring the nuclear reactions from low-energy to
relativistic-energy \cite{PS-PRC-36-2390,KSE-PLB-70-43,PS-PRL-53-1219,SJP-PRL-70-3000,BDH-RMP-62-553,BW-ARNPS-42-77,HU-ARNPS-49-529,WUA-PR-319-145}.
Recently, it has been extended to other fields, for instance, the
analogous correlations describing the fermionic statics of elections \cite{WDO-S-284,HM-S-284}.

In heavy ion collisions (HICs) at intermediate energy, the
HBT method is widely used to extract the space-time properties such as the
source size and emission time of fragments by two-particle
correlation functions \cite{GR-PRL-91-092701,VG-EPJA-30-81}. The dependences
of momentum correlation function on the impact parameter \cite{GWG-PRC-43-781,MYG-PRC-73-014604}, the total momentum of nucleon pairs \cite{CN-PRL-75-4190,MYG-PRC-73-014604}, the isospin
of the emitting source \cite{GR-PRC-69-031605}, the nuclear symmetry
energy \cite{CLW-PRL-90-162701} and the nuclear equation of state
(EOS) \cite{MYG-PRC-73-014604} are also explored by experiment and
theory.

Besides the applications of two-particle momentum correlation
functions to investigate heavy ion collision process, the HBT method is
also extended to study exotic structure of nuclei far from the
$\beta$-stability line due to the rapid development of radioactive
nuclear beams. The neutron-neutron correlation functions (C$_{nn}$)
of Borromean halo nuclei such as $^6$He, $^{11}$Li and $^{14}$Be are
constructed to extract the size of separation between the two
halo neutrons \cite{IK-PRL-70-730,MFM-PLB-476-219,MFM-PRC-64-061301,YMT-PRC-72-01601,PM-PRC-69-011602,PM-NPA-790-235}.
In addition, Wei and Ma \emph{et al}. found that the strength of
proton-neutron momentum correlation functions at small relative
momentum has linear dependence on binding energy per nucleon or single
neutron separation energy for light isotope chains \cite{WYB-PLB-586-225,WYB-JPG-30-2019,MYG-PRC-73-014604}. The
analogous suppressed proton-proton correlation function is also
suggested as another potential tool to diagnose proton halo nuclei \cite{MYG-NPA-790-299} beside the conventional methods such as total reaction cross section and momentum distribution width measurements \cite{TI-PLB-160-380,KT-PRL-60-2599}.
Ma \emph{et al}. have carried our experiment on RIPS in RIKEN to measure the
proton-proton momentum correlation function for revealing the exotic
structure of proton-rich nucleus $^{23}$Al \cite{ZP-IJMPE-19-957,SXY-IJMPE-19-1823}.
Therefore, it is very interesting to investigate how the
exotic structure affects the nucleon-nucleon momentum
correlation function, which can serve as a potential observable to
extract information about anomalous structure in nucleus.

In this paper, we calculated $^{15}$C + $^{12}$C collisions by IDQMD
model. $^{15}$C is a one-neutron halo candidate because of it's small neutron separation energy: $S_{n}$ = 1.218 MeV \cite{AG-NPA-729-337}, closed-neutron shell core: $^{14}$C, narrower momentum distribution of $^{14}$C fragment from the breakup of $^{15}$C \cite{DB-PRC-57-2156,FDQ-PRC-69-034613} and larger $s$-wave spectroscopic factor of $^{15}$C ground state \cite{MG-NPA-579-125} by $^{14}$C($d$,$p$)$^{15}$C reaction measurement. However, a consistent picture have not been obtained in reaction cross section measurement. The interaction cross section ($\sigma_{I}$) do not have peculiarity compared with neighbor isotopes at incident energy 740 MeV/nucleon \cite{OA-NPA-693-32}. However, reaction cross section ($\sigma_{R}$) shows more or less enhancement at intermediate energies and there is also a large difference factor ($d$) for $\sigma_{R}$ \cite{FDQ-PRC-69-034613,FDQ-PRC-61-064311}, which is defined as \cite{OA-NPA-608-63}:
\begin{math}
d = \frac{\sigma_{R}(\exp)-\sigma_{R}(G)}{\sigma_{R}(G)},
\end{math}
where $\sigma_{R}(\exp)$ represents intermediate energy experimental $\sigma_{R}$ and $\sigma_{R}(G)$ is the $\sigma_{R}$ calculated by the Glauber Model at the same bombarding energy with HO-type density distribution, which is obtained by fitting experimental $\sigma_{R}$ at high energy.
The $s$-wave spectroscopic factor extracted from $\sigma_{I}$ is also different from the value obtained from transfer reaction.  Fang \emph{et al}. found that $s$-wave component is dominant in the
ground state of $^{15}$C by simultaneous measurement of $\sigma_{R}$ and momentum distribution \cite{FDQ-PRC-69-034613}. However, the fitted results by the Glauber model are deviated from the experimental data at low energies for
$\sigma_{R}$. Therefore, new probes are needed to estimate the density distribution of valence neutron in $^{15}$C
and other analogous neutron-rich
exotic nucleus candidates.

The initialized $^{15}$C projectiles are sampled from densities calculated by RMF model to study how density distributions of outer neutron affect nucleon-nucleon momentum correlation function.
In RMF model, the last neutron of $^{15}$C is populated on both 1$d$5/2 and 2$s$1/2 states, respectively.
Because the isospin degree of freedom is removed, the relationship between the momentum
correlation function and the structure of exotic nuclei such as skin can be
more directly explored by comparing different collisions induced by different configured $^{15}$C projectiles.

The rest part of this paper is organized as follows: in Sec. \ref{methods},
we briefly describe the HBT technique and the IDQMD model; the initialization of $^{15}$C projectiles and
nucleon-nucleon momentum correlation function of different configured C isotopes induced
collisions are discussed in Sec. \ref{results_and_dis}; the summary is presented in
Sec. \ref{summary}.

\section{HBT technique and IDQMD model}\label{methods}
\subsection{HBT technique}

It is known that the final-state interaction (FSI) and
quantum-statistical symmetry (QSS) affect the wave function of
relative motion of light identical particles when they are emitted in
close region in phase-space and time, which is the principle of
intensity interferometry, i.e., the HBT method. The correlation
function of two-particle can be obtained by convolution of emission
function $g(\textbf{p},x)$ in standard \emph{Koonin-Pratt} equation \cite{PS-PRC-36-2390,KSE-PLB-70-43,PS-PRL-53-1219}:
\begin{equation}
C(\textbf{P},\textbf{q}) = \frac{\int
d^4x_1~d^4x_2~g(\textbf{P}/2,x_1)~g(\textbf{P}/2,x_2)~|\phi(\textbf{q},\textbf{r})|^2}{\int
d^4x_1~g(\textbf{P}/2,x_1)\int d^4x_2~g(\textbf{P}/2,x_2)},
\end{equation}
where \textbf{P}(= \textbf{p}$_1$ + \textbf{p}$_2$) and \textbf{q}[=(
\textbf{p}$_1$ - \textbf{p}$_2$)/2] are the total and relative momentum
of particle pair, respectively,
$g(\textbf{p},x)$ is the probability of emitting a particle with momentum \textbf{p} at space-time point x(\textbf{r}, t) and $\phi(\textbf{q},\textbf{r})$ is two-particle relative wave function with relative distance
\textbf{r} = (\textbf{r}$_2$ - \textbf{r}$_1$) -
$\frac{1}{2}$(\textbf{v}$_1$ + \textbf{v}$_2$)($t_2$ - $t_1$).

In specific application of the Koonin-Pratt formula, the
reliable single-particle phase space distribution at freeze-out
stage is needed \cite{GWG-PRC-43-781}. In this paper, the IDQMD model is used as event
generator. It is a widely used transport model in intermediate
energy HICs and has also been successfully applied to HBT studies for
neutron-rich nuclei induced reactions by Wei and Ma \emph{et al}. \cite{MYG-PRC-73-014604,WYB-PLB-586-225,WYB-JPG-30-2019,MYG-NPA-790-299}. The phase
space of emitted particles is used as the input of Pratt's CRAB (correlation after burner) code \cite{PS-NPA-566-103c}, which takes
the FSI and QSS effects into account for remedying the disadvantage of
semi-classical transport model.

\subsection{IDQMD model}

The quantum molecular dynamics (QMD) approach is a many-body theory
that can describe HICs from intermediate to relativistic energies \cite{AJ-PRL-58-1926,AJ-PR-202-233,PG-PRC-39-1402}.
The main advantage of the QMD model is that it can explicitly treat many-body state of the collision system,
so it contains correlation effects to all orders. Therefore, the QMD model provides valuable
information about both the collision dynamics and the fragmentation
process. The model also has excellent extensibility due to it's
microscopic treatment of collision process. It mainly consists of
several parts: initialization of the projectile and the target
nucleons, propagation of nucleons in the effective potential,
nucleon-nucleon (NN) collisions in a nuclear medium, the Pauli
blocking and the numerical test.

The IDQMD model is based on QMD model and affiliates the isospin
factors in mean field, two-body NN collisions and Pauli blocking \cite{MYG-PRC-73-014604,WYB-PLB-586-225,YTZ-PLB-638-50,CXG-PRC-81-061603,ZGQ-PRC-83-064607,ZGQ-PRC-84-034612}. In
addition, the phase space sampling of neutrons and protons in the
initialization should be separately treated because of the large
difference between neutron and proton density distributions for
nuclei far from the $\beta$-stability line. In order to properly incorporate nuclear structure effects into microscopic simulations, stable initialized $^{15}$C with and without neutron-halo structure have been sampled.

In IDQMD model, each nucleon is represented by a Gaussian wave
packet with width $\sqrt{L}$ (here $L$ = 2.16 $fm^2$) centered
around the mean position $\vec{r_i}(t)$ and the mean momentum
$\vec{p_i}(t)$:

\begin{eqnarray}
\psi_i(\vec{r},t) = \frac{1}{{(2\pi L)}^{3/4}}
\exp[-\frac{{(\vec{r}- \vec{r_i}(t))}^2}{4L}]\exp[\frac{i\vec{r}
\cdot \vec{p_i}(t)}{\hbar}].
\end{eqnarray}
Then all nucleons interact via mean field and NN collisions. The nuclear mean field is parameterized by
\begin{eqnarray}
&&\!\!\!\!
 U(\rho,\tau_{z}) =
\alpha(\frac{\rho}{\rho_{0}}) +\beta(\frac{\rho}{\rho_{0}})^{\gamma}
+\frac{1}{2}(1-\tau_{z})V_{c} \nonumber
\end{eqnarray}
\begin{eqnarray}
&&~~~~~~~+C_{sym}\frac{\rho_{n}-\rho_{p}}{\rho_{0}}\tau_{z} +U^{\rm
{Yuk}},
\end{eqnarray}
with $\rho_{0}$ = 0.16 $fm^{-3}$ (the normal nuclear matter
density). $\rho$, $\rho_{n}$ and $\rho_{p}$ are the total, neutron
and proton densities, respectively. $\tau_{z}$ is $z$th component of
the isospin degree of freedom, which equals 1 or -1 for neutrons or
protons, respectively. The coefficients $\alpha$, $\beta$ and
$\gamma$ are parameters of nuclear equation of state (EOS).
$C_{sym}$ is the symmetry energy strength due to the difference
between neutron and proton, taking the value of 32 MeV. In this work, $\alpha$ = $-$ 356 MeV,
$\beta$ = 303 MeV and $\gamma$ = $7/6$ are taken, which corresponds
to the so-called soft EOS with \cite{AJ-PR-202-233}. $V_{c}$ is the Coulomb
potential and $U^{Yuk}$ is Yukawa (surface) potential, which has the
following form:
\begin{eqnarray}
&&\!\!\!\!
 U^{\rm {Yuk}} =
\frac{V_{y}}{2m}\sum_{{i}\neq{j}}\frac{1}{r_{ij}}{\rm exp}(Lm^{2})\nonumber\\
&&~~~~~~~~\times[{\rm exp}(-mr_{ij}){\rm erfc}(\sqrt{L}m -r_{ij}/\sqrt{4L})\nonumber\\
&&~~~~~~~~-{\rm exp}(mr_{ij}){\rm erfc}(\sqrt{L}m+r_{ij}/\sqrt{4L})],
\end{eqnarray}
with $V_{y}$ = -0.0074 GeV, $m$ = 1.25 $fm^{-1}$ and $r_{ij} = |\vec{r_i}-\vec{r_j}|$ is the relative
distance between two nucleons. Experimental parameterized NN cross
section, which is energy and isospin dependent, is used.

The Pauli blocking effect in IDQMD model is also isospin dependent.
The blocking of neutron and proton is separately treated as follows:
each nucleon occupies a six-dimensional sphere with a volume of
$\hbar^{3}$/2 in the phase space (considering the spin degree of
freedom) and we calculate the phase space volume ($V$) of the
scattered nucleons being occupied by the rest nucleons with the same
isospin as that of the scattered ones. We then compare
2$V$/$\hbar^{3}$ with a random number and decide whether the
collision is blocked or not.

\begin{figure}
\centering
\includegraphics[scale=0.29]{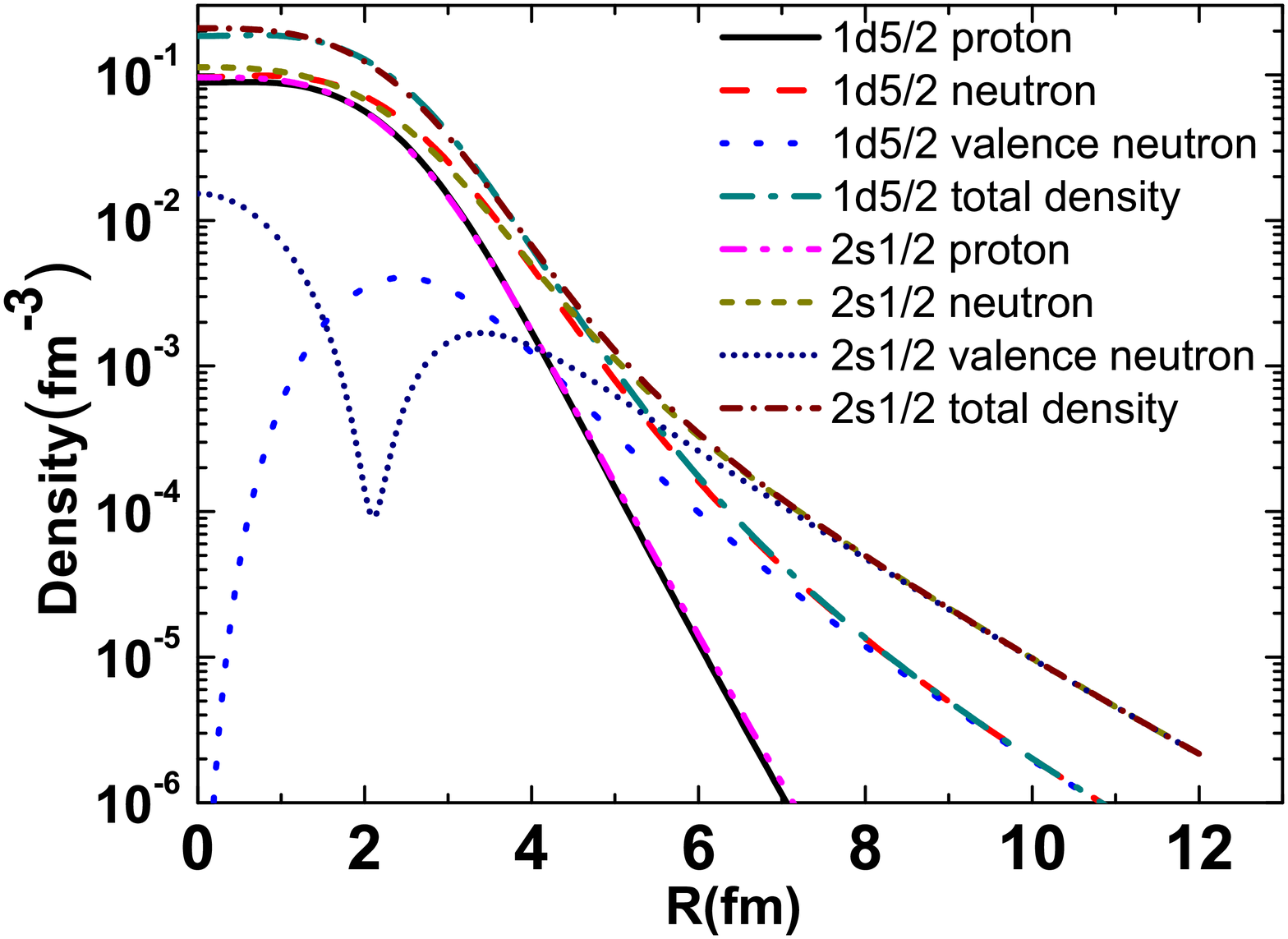}
\vspace{-0.1truein} \caption{\footnotesize (Color online) The proton, neutron, valence neutron and total density distributions of
$^{15}$C for the valence neutron on 1$d$5/2 and 2$s$1/2 states,
respectively, by RMF calculations.}\label{RMF_density2}
\end{figure}

The time evolution of the colliding system is given by the
generalized variational principle. Nuclear clusters are constructed
by a coalescence model, in which particles with relative momentum
smaller than $P_{0}$ and relative distance smaller than $R_{0}$ are
considered to belong to one cluster. The parameter set:
$P_{0}$ = 300 MeV/$c$ and $R_{0}$ = 3.5 $fm$ is taken here.

\section{The initialization of $^{15}$C projectiles and systematical HBT results}\label{results_and_dis}
\begin{table}
\centering
\caption{RMS radii of initialized $^{15}$C by RMF theory and experimental RMS charge radius of $^{14}$C. The RMS radii of $^{15}$C projectiles are initialized to match with the values in this table.}
\label{C15_RMS}
\begin{tabular}{c| c  c  c  c  c  c}
\hline  & Z & N & Valence N & $^{14}$C core & $^{15}$C & $^{14}$C exp. (Z)
\\
  ~~~
&(fm)
&(fm)
&(fm)
&(fm)
&(fm)
&\cite{IA-ADNDT-87-185} (fm)
\\
\hline  &               &      &      &      &      & 2.50\\
\hline  $^{15}$C & 2.40 & 2.76 & 3.85 & 2.51 & 2.62\\
(1$d$5/2)      &               &      &      &      &      &      \\
\hline  $^{15}$C & 2.39 & 2.96 & 5.01 & 2.51 & 2.75\\
(2$s$1/2)      &               &      &      &      &      &      \\
\hline
\end{tabular}
\end{table}

\begin{figure*}
\centering
\includegraphics[width=13.9cm]{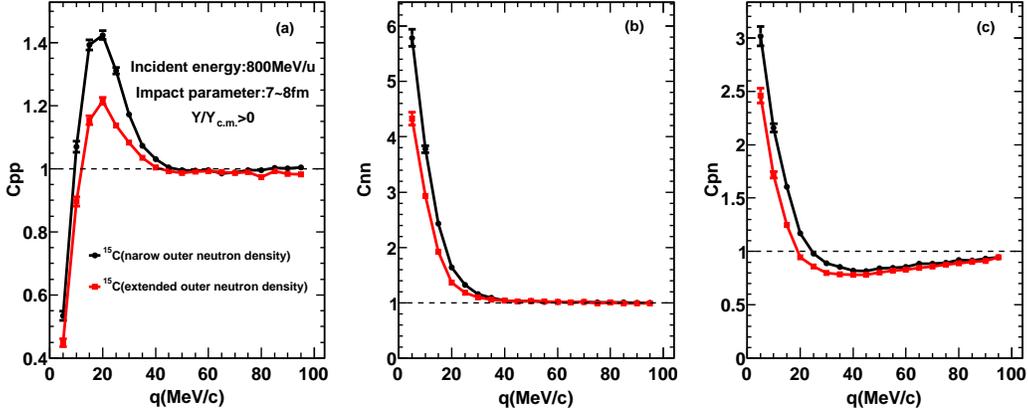}
\vspace{-0.1truein} \caption{\footnotesize (Color online)
Proton-proton ($C_{pp}$, panel (a)), neutron-neutron ($C_{nn}$, panel (b)) and proton-neutron ($C_{pn}$, panel (c)) momentum correlation functions as a
function of relative momentum, which are calculated at incident
energy of 800 MeV/nucleon, impact parameter $b = 7-8 fm$ and selected
nucleons with rapidity $>$ 0. The circles and squares represent two different kinds of initialized $^{15}$C projectiles induced collisions, where $^{15}$C projectiles are sampled from two density outputs of RMF model with valence neutron on
1$d$5/2 and 2$s$1/2 states, respectively.
The dashed line is used to guide the eyes.
}\label{Cpp_Cnn_Cpn_total}
\end{figure*}

\begin{figure}
\centering
\includegraphics[width=8.6cm]{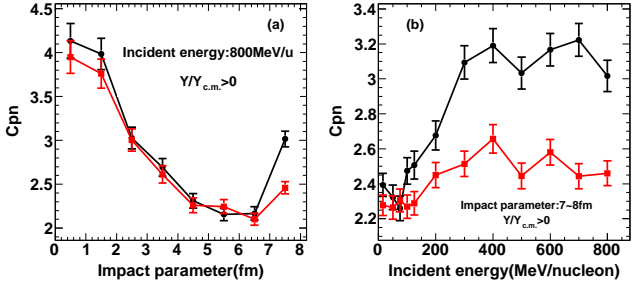}
\vspace{-0.1truein} \caption{\footnotesize (Color online)
(a): $C_{pn}$ strength at
5 MeV/$c$ as a function of impact parameter, which is calculated at
incident energy of 800 MeV/nucleon and selected nucleons with
rapidity $>$ 0; (b): $C_{pn}$ strength at
5 MeV/$c$ as a function of incident energy, which is calculated at
impact parameter $b = 7-8 fm$ and selected nucleons with rapidity
$>$ 0. The symbols in both panel (a) and (b) have the same convention as in Fig. \ref{Cpp_Cnn_Cpn_total}.
}\label{Cpn}
\end{figure}

In IDQMD, the initialization of projectiles and targets distinct proton from neutron.
We sample the nucleon's
coordinates from density distributions of proton and neutron, which
are calculated by relativistic mean field (RMF) method.
Fig. \ref{RMF_density2} shows the proton, neutron, valence neutron and
total density distributions of two different configured $^{15}$C projectiles, whose valence
neutron is assigned on both 1$d$5/2 and 2$s$1/2 states,
respectively. It can be seen that the valence neutron on 2$s$1/2
state has longer tail than that on 1$d$5/2 state while
proton almost has the same density distributions for two case. The
effect of more extended valence neutron density distribution can be
reflected by nucleon-nucleon momentum correlation function as showed
below.

In our calculations, initialization of $^{15}$C is carefully controlled.
The stability of the sampled $^{15}$C projectiles is strictly checked by time
evolution in mean field till 200 $fm/c$ at zero temperature according to the average binding energies,
root-mean-square (RMS) radii and density distributions of neutron and proton.
Eligible initialization samples should meet the following requirements till 200 $fm/c$: a) average binding energy needs to match with experimental data; b) RMS radius also needs to be in according with RMF result; c) the difference of neutron tails between two kinds of $^{15}$C projectiles should keep all the way.
To better imitate input density from RMF calculation and reflect the structure effect, thousands of eligible initialized samples are accumulated to simulating collisions.
The initialized samples of other C isotopes are prepared in a similar way above.

Even with very elaborate initialization, the tail of density in QMD model can not be reproduced very well compared with RMF model since the wave function in QMD model has Gaussian form. However, the obvious differences of neutron tails between two kinds of $^{15}$C projectiles can keep stable enough and thus it can play the role of skin structure in collisions. In the following, we will present how the skin structure expressed by different neutron density distributions in QMD affects nucleon-nucleon momentum correlations.

The proton and neutron phase space of $^{15}$C + $^{12}$C collisions at freeze-out time generated by IDQMD model, is used as input to CRAB code. The obtained proton-proton, neutron-neutron and proton-neutron momentum correlation functions are shown in
Fig. \ref{Cpp_Cnn_Cpn_total}, respectively, where $q$ denotes relative momentum of
nucleon pair. As expected, $C_{nn}$ and $C_{pn}$ peak at small $q$ while proton-proton is anti-correlation at small $q$ owning to the Coulomb potential and antisymmetrization. The peak of $C_{pp}$ at 20 MeV/$c$ is due to $s$-partial wave of the proton-proton
scattering, which strongly depends on the size of emitting source. Our simulations pretty well reproduce the shape and height of $C_{pp}$, $C_{nn}$ and $C_{pn}$ vs. $q$ compared with experimental cases \cite{VG-EPJA-30-81}.
It can be seen that $C_{nn}$ and $C_{pn}$ are both largely reduced at small $q$ because of  the more extended neutron density distribution.
In Fig. \ref{Cpp_Cnn_Cpn_total},
the size of emission source is mostly decided by projectile-like remnants since
we adopt impact parameter $b = 7 - 8 fm$ and rapidity $>$ 0.
The RMS radii of proton, neutron, valence neutron, $^{14}$C core, $^{15}$C by RMF calculation and experimental RMS charge radius of $^{14}$C are shown in table \ref{C15_RMS}. The RMS radii of initialized $^{15}$C projectiles are required to meet the RMF calculated values.
Though there is no available experimental RMS radius of $^{15}$C, RMS radius of $^{14}$C core reproduces experimental RMS charge radius of $^{14}$C quite well.
The RMS radii of $^{15}$C do not have large difference for two kinds of $^{15}$C projectiles.
However, the corresponding RMS radii of valence neutron are 3.85 $fm$ and 5.01 $fm$.
Therefore, the strength of momentum correlation functions can indeed sensitively
reflect the fine difference of neutron density distribution.
Although the density distributions of proton for two configured $^{15}$C are almost the same (as seen in Fig. \ref{RMF_density2}),
$C_{pp}$ still has large differences, which demonstrate that the proton density distribution has changed when proton emission. The density distributions of neutrons and protons are coupling with each other during collisions.

\begin{figure}
\centering
\includegraphics[width=8.6cm]{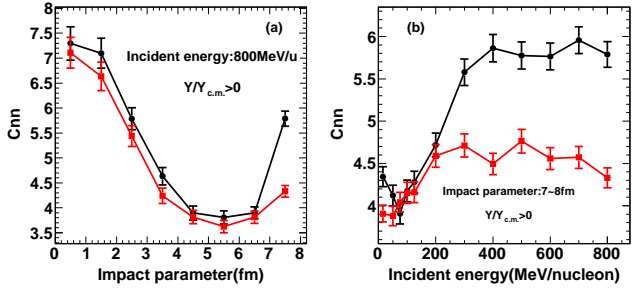}
\vspace{-0.1truein} \caption{\footnotesize (Color online)
The same as Fig. \ref{Cpn} but for $C_{nn}$.}\label{Cnn}
\end{figure}

The impact parameter dependence of $C_{pn}$ at 800 MeV/nucleon incident energy for the two different kinds of
initialized $^{15}$C projectiles is shown in panel (a) of
Fig. \ref{Cpn}. The strength of correlation becomes
decreasing with b up to 6.5 $fm$. In central collisions, emitted
nucleons have stronger correlations between themselves because they come from
one compact, hot and dense region, which is consistent with previous
Boltzmann-Uehling-Uhlenbeck (BUU) \cite{GWG-PRC-43-781,CLW-PRC-68-014605} and IDQMD \cite{MYG-PRC-73-014604} results. In peripheral
collisions with $b > 6.5 fm$ the rapid enhancement of $C_{pn}$
reveals different collisions dynamics from central collisions. This is due to the fact that there is
no bulk overlap region at large impact parameter and only the
outer nucleons of $^{15}$C are scraped. They keep more initial structure information of projectile. Thus, there exists strong correlation among these nucleons.
The difference of $^{15}$C in neutron density distribution such as skin structure can be well revealed in peripheral collisions while the small difference of density is wiped away in violent central and semi-peripheral collisions.

How incident energy affects $C_{pn}$ is represented in panel (b) of Fig. \ref{Cpn}. The correlations for both cases increase from very low incident energy and then almost reach saturation above
300 MeV/nucleon.
This is understandable due to the fact that increasing
incident energies lead to more rapid collision process and there is smaller space and time interval among emitting nucleons \cite{CLW-PRC-68-014605,MYG-PRC-73-014604}, which leads to stronger correlation.
The apparent differences in $C_{pn}$ exist for the two kinds of $^{15}$C induced collisions above saturation energy.
The saturation of $C_{pn}$ at high incident
energies and the large differences provide us the proper entrance channel conditions to explore the relation
between initial structure and final effect.

\begin{figure}
\centering
\includegraphics[width=8.6cm]{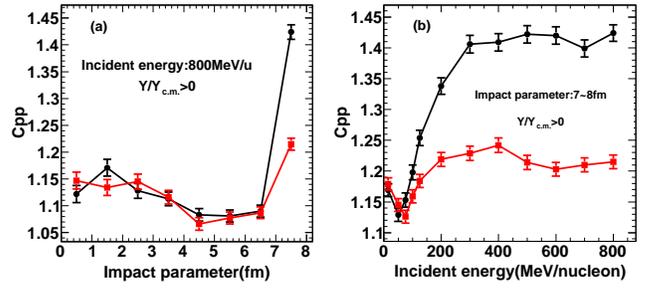}
\vspace{-0.1truein} \caption{\footnotesize (Color online)
The same as Fig. \ref{Cpn} but for $C_{pp}$.}\label{Cpp}
\end{figure}

The impact parameter and incident energy dependences of $C_{nn}$ are shown in panel (a) and panel (b) of Fig. \ref{Cnn}, respectively. The tendency of $C_{nn}$ is similar to $C_{pn}$ for both impact parameter and energy dependences.
The analogous results are also obtained for $C_{pp}$ shown in Fig. \ref{Cpp}.

\begin{figure}
\centering
\includegraphics[width=8.6cm]{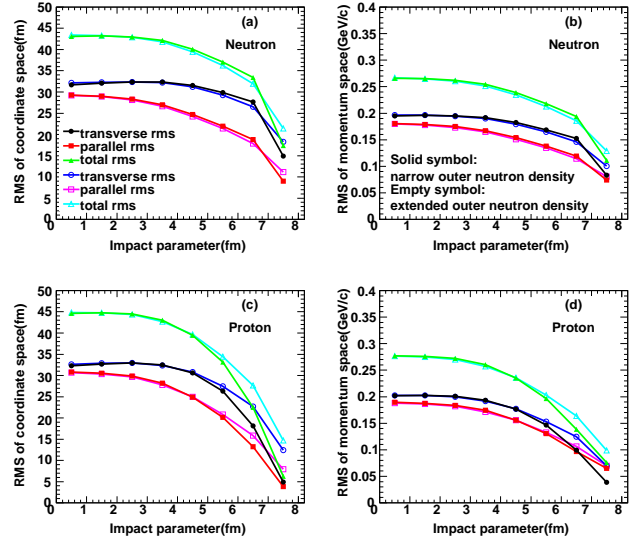}
\vspace{-0.1truein} \caption{\footnotesize (Color online)
Root-mean-square of coordinate (panel (a) and (c)) and momentum (panel (b) and (d)) space distributions in the
center of mass reference frame as a function of impact parameter at incident
energy of 800 MeV/nucleon and rapidity $>$ 0 for neutron (panel (a) and (b)) and proton (panel (c) and (d)), respectively. The solid (empty) circles, squares and triangles represent transverse, parallel and total RMS radii, respectively, for $^{15}$C projectiles induced collisions, where $^{15}$C projectiles are sampled from two density outputs of RMF model with valence neutron on
1$d$5/2 and 2$s$1/2 states, respectively.
}\label{12_ene_imp_dependence}
\end{figure}

\begin{figure}
\centering
\includegraphics[width=8.6cm]{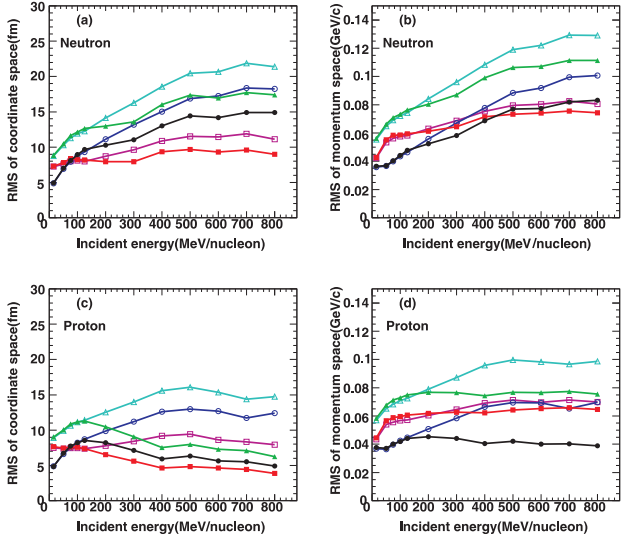}
\vspace{-0.1truein} \caption{\footnotesize (Color online)
Root-mean-square of coordinate (panel (a) and (c)) and momentum (panel (b) and (d)) space distributions in the
center of mass reference frame as a function of incident energy with impact parameter $b = 7-8 fm$ and rapidity $>$ 0 for neutron (panel (a) and (b)) and proton (panel (c) and (d)), respectively. The symbols have the same convention as in Fig. \ref{12_ene_imp_dependence}.}\label{8_b_7-8fm_ene_dependence}
\end{figure}

The strength of $C_{pn}$, $C_{nn}$ and $C_{pp}$ is mainly decided by the distance in phase space at freeze-out time. Under specific entrance channel conditions (high energy and large impact parameter), the size of phase space in final state
is consistent with initial size of $^{15}$C projectile, which largely depends on the neutron skin structure.
This can be seen from Fig. \ref{12_ene_imp_dependence} showing impact parameter dependences of RMS radii in coordinate and momentum space. The size of phase space decreases with the increasing of impact parameter.
The more extended neutron skin leads to larger RMS radii of coordinate and momentum in final state for peripheral collisions at 800 MeV/nucleon, which results in smaller correlation in the HBT method. However, RMS radii do not have difference between the two kinds of $^{15}$C induced reactions in central and semi-peripheral collisions because the only difference in outer neutron is covered in violent collisions. Therefore, the physical picture of phase space size is consistent with impact parameter dependences of $C_{pn}$, $C_{nn}$ and $C_{pp}$.
The excitation function of RMS radii of phase space in peripheral collisions shown in Fig. \ref{8_b_7-8fm_ene_dependence} gradually become saturated at high incident energy, which can explain the saturation of correlations in panel (b) of Fig. \ref{Cpn}, \ref{Cnn} and \ref{Cpp}.

\begin{figure}
\centering
\includegraphics[width=8.6cm]{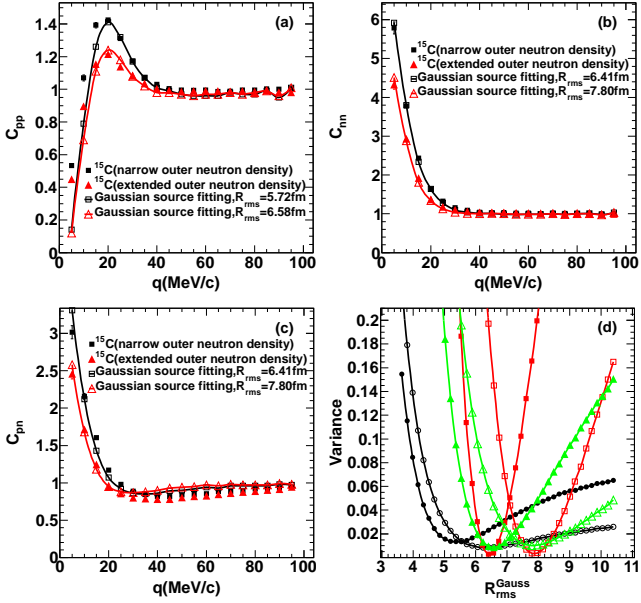}
\vspace{-0.1truein} \caption{\footnotesize (Color online)
Gaussian source fitting for $C_{pp}$ (panel (a)), $C_{nn}$ (panel (b)), $C_{pn}$ (panel (d)) and variance between IDQMD + CRAB correlation and Gaussian source correlation as a function of Gaussian source size (panel (d)).
The entrance channel conditions of 800 MeV/nucleon bombarding energy and impact parameter with $b = 7-8 fm$ are used and the nucleons with rapidity $>$ 0 in final state are chosen.}\label{Gauss_source_fit}
 \end{figure}

\begin{figure}
\centering
\includegraphics[width=8.6cm]{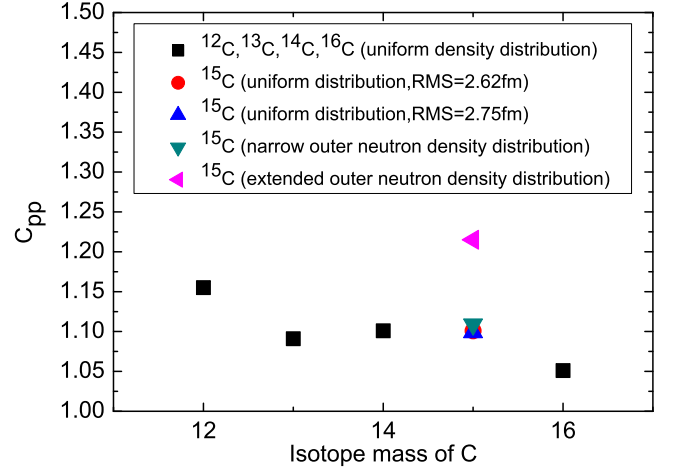}
\vspace{-0.1truein} \caption{\footnotesize (Color online)
Dependence of $C_{pp}$ (strength of $C_{pp}$ at 20 MeV/$c$) on C isotopes ($^{12,13,14,15,16}$C) induced collisions. Different collision systems are compared with the same reduced impact parameter range: 0.875~--~1.0 and only protons with rapidity $>$ 0 are used to calculate momentum correlation function. Nucleons of $^{12,13,14,16}$C are sampled from uniform density. We construct four kinds of $^{15}$C projectiles: two of them are sampled from two RMF densities shown in Fig. \ref{RMF_density2}, respectively, and the other two are sampled from uniform distribution with two different sizes, corresponding to the sizes of 1$d$5/2 and 2$s$1/2 states in RMF model, respectively.
}\label{sys_dep_reduced_impact}
\end{figure}

The space-time extent of emission source for different particles can be
extracted from the shape and height of $C_{pp}$, $C_{nn}$ and $C_{pn}$ for $^{15}$C projectiles with and without neutron skin.
Angle-averaged Koonin-Pratt formula can be written as \cite{KSE-PLB-70-43,VG-PRC-65-054609}:
\begin{equation}
R(q) = 4\pi\int r^2drK(q,r)S(r),
\end{equation}
where $S(r)$ is the isotropic source function standing for the probability distribution of emitting a pair of nucleons at relative distance $r$. $K(q,r)$ is angle-averaged kernel function obtained from radial part of two-nucleon relative wave function. If we assume $S(r)$ has simple Gaussian form:
\begin{equation}
S(r) = \frac{1}{(2\pi)^{3/2}\sigma^3}\exp(-\frac{r^2}{2\sigma^2}),
\end{equation}
where $\sigma$ describes the spatial distribution of nucleons emitting location and the RMS radius of Gaussian source equals $\sqrt{3}\sigma$. Gaussian source method is used to fit the correlation functions ($C_{pp}$, $C_{nn}$ and $C_{pn}$) for two kinds of $^{15}$C induced collisions.
The panel (d) of Fig. \ref{Gauss_source_fit} represents the variance between IDQMD and Gaussian source correlations as a function of RMS radius of Gaussian source.
Then the best fitting is judged by the variance and the RMS radii obtained are shown in panel (a), (b) and (c), respectively.
We get the same Gaussian source size from $C_{nn}$ and $C_{pn}$. The emission source size of projectiles with skin structure is about 20\% larger than that of projectiles without skin structure for $C_{nn}$ and $C_{pn}$.
The source size of $C_{pp}$ is smaller than that of $C_{nn}$ and $C_{pn}$ for both kinds of $^{15}$C projectiles.
Comparing with initialized neutron distribution shown in Fig. \ref{RMF_density2} and table \ref{C15_RMS}, we can conclude that the initial structure information of neutron skin and size of projectile is kept till final state.
The size of proton emission source is different for the two kinds of $^{15}$C projectiles, though initial density distribution of proton is the same.
The size is about 15\% larger for projectiles with neutron skin, which proves that proton density distribution is disturbed through interaction with neutron even in very peripheral collisions.

The size of emission source extracted above is under hypothesis that nucleon is expressed by point particle. However, in QMD model, nucleon wave function has Gaussian form in coordinate and momentum space. The true distribution of source is the convolution of the distribution of wave packet center with Gaussian density distribution of single nucleon. In Ref. \cite{JA-NPA-617-510,FG-PRC-65-014901}, the relation between RMS radius of the freeze-out points $<r(t)^2>^{1/2}$ and the variance of the source and wave function is obtained by assuming that a chaotic Gaussian source has formed in collision without correlation between coordinate and momentum space:
\begin{equation}
<r(t)>^{1/2} = \sqrt{3(A+L)},
\end{equation}
where $A$ is variance of chaotic emission source with Gaussian form and $L$ is square of width of Gaussian wave packet in QMD model with value of 2.16 $fm^2$ in our calculations. Now with finite width modification of Gaussian wave packet, then we get smaller RMS radius of emission source: $\sqrt{3A} = \sqrt{3(\sigma^2 -L)}$. For example, RMS radii of $C_{pp}$ are 5.12 $fm$ and 6.07 $fm$ for $^{15}$C projectiles with and without neutron skin, respectively. RMS radii of both $C_{nn}$ and $C_{nn}$ have the same value: 5.88 $fm$ and 7.37 $fm$, respectively.

We also study the systematical dependence of momentum correlation on other C isotopes induced collisions. Fig.~\ref{sys_dep_reduced_impact} shows the strength of $C_{pp}$ at 20 MeV/$c$ for $^{12,13,14,15,16}$C + $^{12}$C systems.
The collisions are compared at the same reduced parameter range: 0.875~--~1.0. Four kinds of $^{15}$C projectile are constructed. The other two kinds of $^{15}$C projectiles are sampled from uniform density distributions beside the sampling from RMF discussed above. Nucleons are also uniform distribution in $^{12}$C, $^{13}$C, $^{14}$C and $^{16}$C projectiles.
The total systematical tendency of mass dependence of $C_{pp}$ is decreasing except for projectiles with neutron skin structure, which can be interpreted as following: though they are simulated at the same reduced impact parameter, $^{12}$C + $^{12}$C collisions have smaller emission source as compared with $^{16}$C + $^{12}$C collisions.
Correlation for projectiles with neutron skin is stronger than other cases. This is because that the outer density distribution is more extended due to neutron skin and then it results in smaller overlap zone between projectile and target for the same reduced impact parameter at peripheral collisions.
Therefore, proton-proton has stronger correlation due to they come from one compact projectile-like remnants, which is obvious different from the case without skin structure.

\section{Summary}\label{summary}
The very small nucleon separation energy of last nucleon and suddenly increased radius compared with its neighboring isotopes are two main features of exotic halo nucleus. HBT method has been used to study the relationship between momentum correlation and single nucleon separation energy several years before.
In this paper, we systematically investigate how the density distribution of valence neutron in one-neutron halo candidate $^{15}$C affects the strength of nucleon-nucleon momentum correlation function.
Specifically, two kinds of $^{15}$C projectiles are sampled from two different density outputs of RMF model, respectively. The difference of outer neutron density distribution between two kinds of initialized $^{15}$C samples can be viewed as neutron skin structure in IDQMD simulation.
The more extended density distribution due to outer neutron in projectile results in the larger and hot emission source, which leads to a larger size of phase space in final state and then corresponds to weaker correlation.
The energy and impact parameter dependences show that
nucleon-nucleon momentum correlation function is a very sensitive observable of density distribution of valence neutron at high energy and peripheral collisions.
Therefore, momentum correlation function at high bombarding energy and large impact parameter can serve as a new potential probe to diagnose the exotic structure such as skin and halo besides the traditional
measurements of total reaction cross section and momentum distribution.

\section{Acknowledgments}
We wish to thank Prof. S. Pratt for providing CRAB code which is used
to construct the momentum correlation function from
phase space data.
This work is partially supported by National Natural Science
Foundation of China under contract No.s 11035009, 11005140, 10979074, 10875160, 10805067 and 10975174,
 and the Knowledge Innovation Project of the
Chinese Academy of Sciences under Grant No. KJCX2-EW-N01.

\end{CJK*}
\end{document}